\documentclass{article}
\usepackage{dcase2022,amsmath,graphicx,url,times,booktabs, tabularx}
\usepackage{color,makecell,bbm,amssymb,multirow,rotating,diagbox}
\usepackage{utfsym}

\usepackage{cite}
\usepackage{bbding}
\usepackage{balance}
\usepackage{color,makecell,url,times, tabularx,bbm,amssymb,multirow,rotating,diagbox}
\usepackage{algorithmic,algorithm}

\title{A Mixed Supervised Learning Framework for Target Sound Detection}
\name{Dongchao Yang$^{1}$, Helin Wang$^{1}$, Wenwu Wang$^{2}$, Yuexian Zou$^{1}$$^{*}$\thanks{$^{*}$ Corresponding Author: zouyx@pku.edu.cn}\thanks{This paper was partially supported by Shenzhen Science \& Technology Research Program (No: GXWD20201231165807007-20200814115301001; No: JSGG20191129105421211) and NSFC (No: 62176008).}}
\address{$^1$ADSPLAB, School of ECE, Peking University, Shenzhen, China\\
$^2$Center for Vision, Speech and Signal Processing, University of Surrey, UK}
\begin{document}

\ninept
\maketitle

\begin{sloppy}

\begin{abstract}
Target sound detection (TSD) aims to detect the target sound from mixture audio given the reference information. Previous works have shown that TSD models can be trained on fully-annotated (frame-level label) or weakly-annotated (clip-level label) data. However, there are some clear evidences show that the performance of the model trained on weakly-annotated data is worse than that trained on fully-annotated data. To fill this gap, we provide a mixed supervision perspective, in which learning novel categories (target domain) using weak annotations with the help of full annotations of existing base categories (source domain). To realize this, a mixed supervised learning framework is proposed, which contains two mutually-helping student models (\textit{f\_student} and \textit{w\_student}) that learn from fully-annotated and weakly-annotated data, respectively. The motivation is that  \textit{f\_student} learned from fully-annotated data has a better ability to capture detailed information than \textit{w\_student}. Thus, we first let \textit{f\_student} guide \textit{w\_student} to learn the ability to capture details, so \textit{w\_student} can perform better in the target domain. Then we let \textit{w\_student} guide \textit{f\_student} to fine-tune on the target domain. The process can be repeated several times so that the two students perform very well in the target domain. To evaluate our method, we built three TSD datasets based on UrbanSound and Audioset. Experimental results show that our methods offer about 8\% improvement in event-based F-score as compared with a recent baseline.
\end{abstract}

\begin{keywords}
 Target sound detection, audioset, weakly supervised, mixed supervised learning
\end{keywords}

\section{Introduction}
In target sound detection (TSD) \cite{yang2021detect}, one aims to recognize and localize the target sound source within a mixture audio given a reference audio, e.g. detecting the dog bark sound within a street. TSD can be applied to numerous potential fields\cite{bello2018sonyc, stowell2015acoustic, hershey2017cnn}, 
such as species migration monitoring and large-scale multimedia indexing. Sound event detection (SED) \cite{kong2019sound} is a similar task with TSD, and a lot of works have been done for SED \cite{dinkel2021towards,lin2020specialized,kong2020sound,mesaros2021sound, wang2019comparison,martin2019sound}. However, SED aims to classify and localize all pre-defined events (\textit{e.g.}, dog barking, man speaking) within an audio clip, which significantly limits the flexibility to detect unseen classes. Different with SED, TSD only focuses on detecting the event that we interest. TSD does not require pre-defined set of categories, therefore, it can be easily extended to sound detection from an open set.
\begin{figure}[t]
  \centering
  \includegraphics[width=\linewidth]{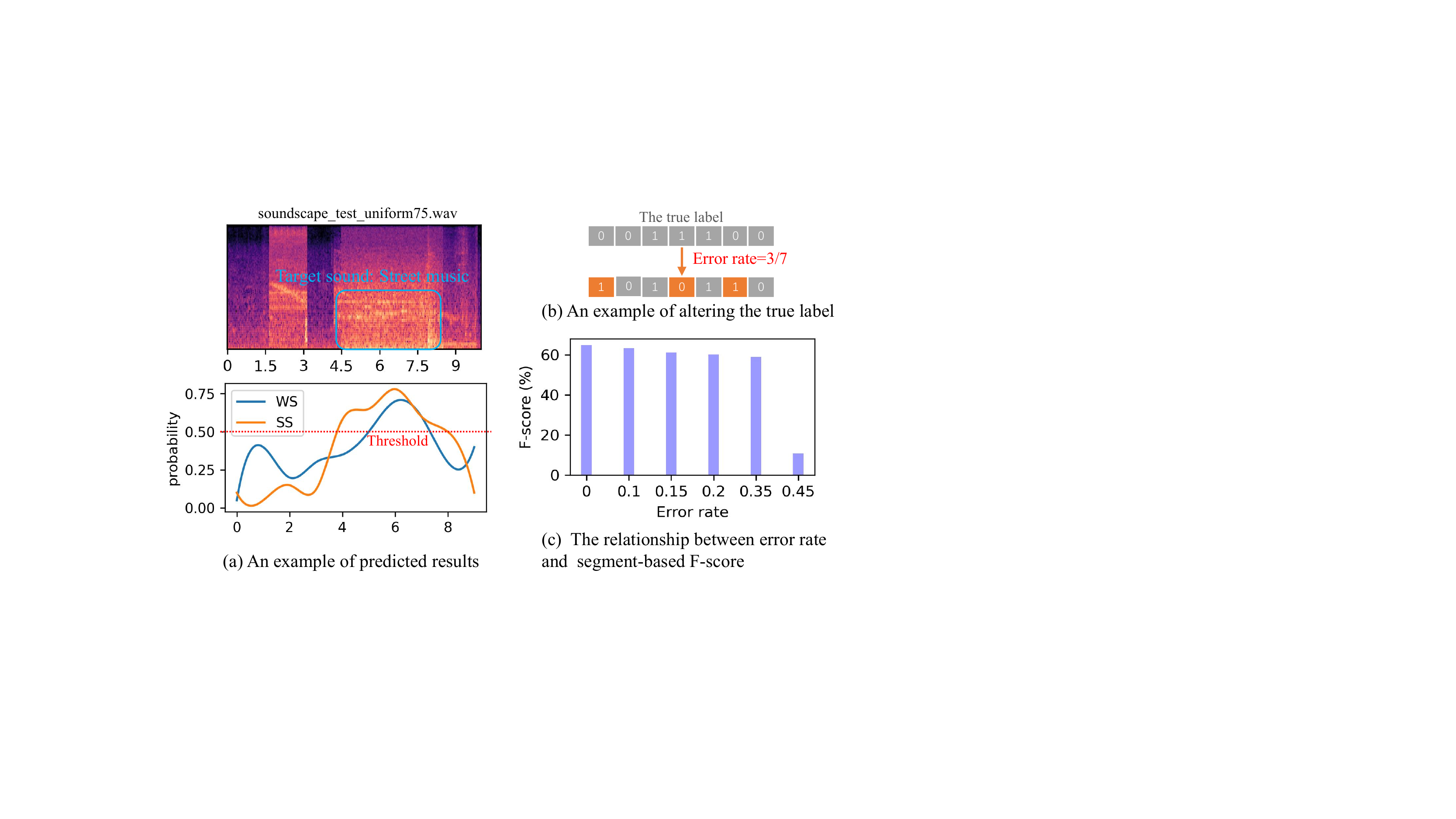}
  \caption{(a) shows the predicted results generated by weakly (WS) and strongly supervised (SS) learning. (c) shows the influence of the error rate of the frame-level label when we train TSDNet model \cite{yang2021detect} on the URBAN-TSD dataset.}
  \label{fig:mask}
\end{figure}
In a recent work, a target sound detection network (TSDNet) \cite{yang2021detect} is presented, where a conditional network is used to generate sound-discriminative embedding which is then used as the reference information to guide a detection network for the detection of the target sound from the mixture audio. TSDNet provides a good detection performance when training data is fully-annotated, e.g. the onset and offset time of the target sound are provided in the annotations. However, collecting large-scale fully-annotated data is time-consuming and labor-intensive. Weakly supervised TSD is an effective method to reduce the reliance on fully-annotated data, but the performance tends to degrade significantly \cite{yang2021detect}. 

In this paper, we consider TSD with mixed supervision, which learns novel sound categories (target domain) using weak annotations with the help of full annotations of the existing base sound categories (source domain). Under this setting, we can use a small-scale fully-annotated dataset (e.g. URBAN-SED \cite{salamon2017scaper}) to complement a large-scale weakly-annotated dataset (e.g. Audioset \cite{gemmeke2017audio}). To achieve this, we propose a novel mixed supervised learning framework, which includes two mutually-helping student models (\textit{f\_student} and \textit{w\_student}), which are trained by fully- and weakly-annotated data, respectively. The proposed method involves three novel aspects. Firstly, the \textit{f\_student} learned from fully-annotated data has better ability in capturing detailed information than the \textit{w\_student}. As Figure 1 (a) shows, the model trained on weakly-annotated data fails to locate the event boundary: it only focuses on the most distinct part and misses the boundary information. Thus, we propose a frame-level knowledge distillation (KD) strategy to transfer the knowledge from \textit{f\_student} to \textit{w\_student}, which makes \textit{w\_student} able to capture more details, \textit{e.g.} boundary information. However, it is hard to transfer all of the knowledge from \textit{f\_student} to \textit{w\_student}. Thus, we propose to directly apply  \textit{f\_student} to the target domain with the guidance of  \textit{w\_student}. Specifically, \textit{w\_student} is used to produce frame-level pseudo labels for \textit{f\_student}. This strategy inspired by an interesting phenomenon that even if there are some errors in the frame-level labels, the detection performance remains stable or decreases only slightly (when the error rate lower than 0.35), as Figure 1 (c) shows. The process of mutually helping can be repeated several times so that the two students perform very well in the target domain. Lastly, we found that the mismatch between source and target data distribution tends to affect significantly the performance of transfer learning, \textit{e.g.} URBAN-SED \cite{salamon2017scaper} and Audioset \cite{hershey2021benefit}. Thus, we propose an adversarial training strategy to solve the domain mismatch problem. To evaluate our method, we built two small-scale fully-annotated datasets and a large-scale weakly-annotated dataset based on URBAN-SED and Audioset. Experimental results show that two small-scale fully annotated datasets could significantly improve the performance on large-scale weakly-annotated dataset.
\section{Related Work}
Many methods \cite{kumar2017audio,liang2021joint, lin2020guided, shi2019hodgepodge,chan2020non,zheng2021improved,cances2021comparison,guan2022sparse} have been proposed to utilize both fully- and weakly-annotated data to train the SED model. However, previous methods assume that fully- and weakly- annotated data belong to the same set of pre-defined categories. Our method aims to use the existing base categories with full annotations to facilitate the recognition of novel categories with weak labels. This setting is more realistic for the reason that a small-scale fully-annotated dataset with few categories is, in practice, easier to create, as compared with a large-scale fully-annotated dataset of many categories. 
\section{TSD Datasets} \label{audio-tsd}
We built two TSD datasets based on Audioset \cite{hershey2021benefit}, which includes 94126 training clips and 16118 test clips, from 456 different classes. One is the large-scaled TSD dataset, named as L-TSD dataset, by choosing 192 different classes from Audioset. The other is the small-scaled fully-annotated TSD dataset (S-TSD), by choosing 51 different classes from Audioset. L-TSD includes two types of annotated samples, that is, fully-annotated (\textit{i.e.} L-TSD-strong) and weakly-annotated (\textit{i.e.} L-TSD-weak) samples. There is no common class between L-TSD and S-TSD datasets. The process of building the datasets is similar to the one described in \cite{yang2021detect}. The mixture audio signals come from Audioset. If there are $N$ sound events in the mixture audio, we can generate $N$ positive samples. We also generate $N/2$ negative samples, which do not contain the target sound.
To prepare reference audio, we select the audio clips for each category directly from the Audioset training set i.e. those that do not contain interference from other events. In total, L-TSD includes 490,336 training and 83,334 test clips, while S-TSD contains 26,247 training and 5113 test clips. In addition, we choose the URBAN-TSD dataset \cite{yang2021detect} as another small-scale fully-annotated dataset.
\begin{figure}[t]
  \centering
  \includegraphics[width=\linewidth]{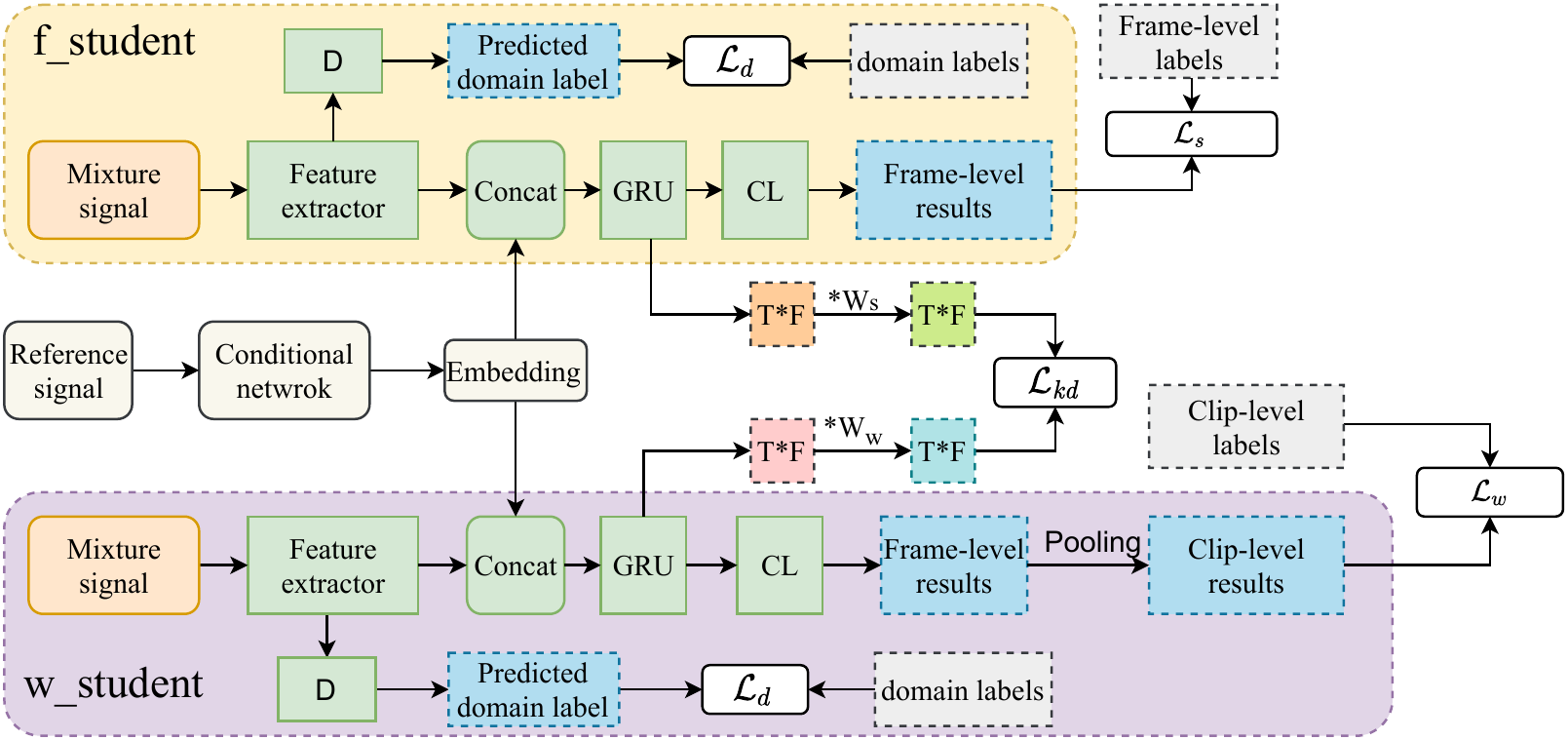}
  \caption{The architecture of the mixed supervised learning framework. CL denotes classification layers, which includes two fully-connected layers and one softmax function. D denotes the discriminator, which consists of three convolutional layers and one fully-connected layer.}
  \label{fig:two_student}
\end{figure}
\section{Proposed Method}
Figure \ref{fig:two_student} shows our proposed mixed supervised learning framework. The core idea of the framework is that the two students can teach each other iteratively. One of the student models is trained on fully-annotated data, we name it as \textit{f\_student}. The other model is trained on weakly-annotated data, and we name it as \textit{w\_student}. The two students have the same structure, while the only difference is that \textit{w\_student} has a linear softmax pooling layer \cite{wang2019comparison}.
\subsection{Network Structure} \label{network structure}
\noindent \textbf{Conditional network.} The conditional network aims to extract a sound-discriminative
embedding vector from the reference audio. Similar to the previous work \cite{yang2021detect}, we adopt a VGG-like convolutional neural network (CNN) model \cite{kong2020panns} for the conditional network.\\ 
\textbf{Detection network.} Similar to the previous work \cite{yang2021detect}, the network is composed of 5 convolutional layers, 1 Bi-GRU layer, and 2 fully-connected layers.
Given the mel-spectrogram of the mixture audio $\boldsymbol{x} \in \mathcal{R}^{T \times F}$, where $T$ and $F$ denote the number of frames and the dimension of frame. The detection network aims to predict frame-level probabilities 
\begin{equation}\label{flp}
\setlength{\abovedisplayskip}{4pt}
\setlength{\belowdisplayskip}{4pt}
\begin{split}
  \hat{p}_i = \mathbb{P}(Y=k|X=x_i,\boldsymbol{e};\phi)
  \end{split}
\end{equation}
where $\phi$ denotes the trainable parameters of the detection network, $\boldsymbol{e}$ denotes the embedding obtained from the conditional network and $x_i$ denotes the $i$-th frame of the mixture audio $\boldsymbol{x}$. For \textit{f\_student}, given the ground-truth label $p_i \in \{0,1\}$ for each frame, which can be optimized by minimizing the binary cross entropy (BCE) loss:
\begin{equation}\label{ssn}
\setlength{\abovedisplayskip}{4pt}
\setlength{\belowdisplayskip}{4pt}
\begin{split}
  \mathcal{L}_{s} = \sum_{i=1}^{t}(- p_i\log\hat{p}_i-(1-p_i)\log(1-\hat{p}_i))
  \end{split}
\end{equation}
where $t$ indicates the number of frames. The difference between \textit{f\_student} and \textit{w\_student} is that the latter needs a pooling layer to get the clip-level prediction. Thus, a LinSoft pooling layer \cite{wang2019comparison} is added after the last layer of the \textit{f\_student}. \textit{w\_student} aims to predict a clip-level probability $\hat{P} = f_{LSP}(\hat{p}_1,\hat{p}_2, ..., \hat{p}_t)$
where $f_{LSP}(\cdot)$ denotes the LinSoft pooling function.
Given the clip-level ground-truth label $P \in \{0,1\}$, the BCE loss is applied as the loss function:
\begin{equation}\label{wsn}
\setlength{\abovedisplayskip}{4pt}
\setlength{\belowdisplayskip}{4pt}
\begin{split}
   \mathcal{L}_{w} = -P\log\hat{P}-(1-P)\log(1-\hat{P})
   \end{split}
\end{equation}
\subsection{Two-student Learning}
In this part, we introduce the details of how to enable \textit{w\_student} and \textit{f\_student} to help each other.\\
\textbf{Frame-level knowledge distillation.}
According to formula (\ref{wsn}), we can see that \textit{w\_student} makes a decision on the whole audio clip. Compared to \textit{f\_student}, \textit{w\_student} is limited in capturing the detailed information of the sound events.
To address this issue, we propose to first train \textit{f\_student} on a small-scale fully-annotated dataset (\textit{i.e.} source domain), and then transfer its knowledge to \textit{w\_student}, so that \textit{w\_student} can get better performance on weakly-annotated dataset (\textit{i.e.} target domain). Specifically, we first train the \textit{f\_student} model on the source dataset with strong labels. After that, we train the \textit{w\_student} model on the source data with weak labels. We then treat the trained \textit{f\_student} model as a teacher, to generate a frame-level feature representation. As a result, \textit{w\_student} may capture more detailed information, due to the frame-level class-agnostic knowledge distillation. More specifically, we can train \textit{w\_student} with the following objective function,
\begin{equation}\label{Mul-task loss}
\setlength{\abovedisplayskip}{4pt}
\setlength{\belowdisplayskip}{4pt}
\begin{split}
   \mathcal{L}_{\mathit{w\_kd}} = \mathcal{L}_{\mathit{w}} + \mathcal{L}_{\mathit{kd}}
   \end{split}
\end{equation}
\begin{equation}\label{kd_loss}
    \mathcal{L}_{\mathit{kd}} = ||\boldsymbol{F}_s \cdot  \boldsymbol{W}_s-\boldsymbol{F}_w \cdot \boldsymbol{W}_w||_2
\end{equation}
where $\boldsymbol{F}_s$ and $\boldsymbol{F}_w$ denote the feature map of the GRU layer of the two models, and $\boldsymbol{W}_s$ and $\boldsymbol{W}_w$ denote the transformation matrix.\\
\textbf{Pseudo Supervised Training.}
The idea of pseudo supervised training strategy is motivated by an interesting observation, i.e. even if there are some errors in the frame-level labels, the detection performance remains stable or decreases only slightly. This means that we could use the noisy frame-level labels as our training target. In this paper, we propose to use \textit{w\_student} to produce noisy frame-level labels (i.e. pseudo labels), and then use the pseudo labels to re-train \textit{f\_student}, as follows
\begin{equation}\label{pseudo1}
    \hat{p}_i^w = \mathbb{P}(Y=k|X=x_i,\boldsymbol{e};\phi_w),
    \hat{p}_i^s = \mathbb{P}(Y=k|X=x_i,\boldsymbol{e};\phi_s)
\end{equation}
\begin{equation}\label{pseudo3}
\setlength{\abovedisplayskip}{4pt}
\setlength{\belowdisplayskip}{4pt}
\begin{split}
   \mathcal{L}_{re\_s} = \sum_{i=1}^{t}(-\hat{p}^w_{i}\log\hat{p}^s_{i}-(1-\hat{p}^w_i)\log(1-\hat{p}^s_i))
   \end{split}
\end{equation}
\noindent \textbf{Adversarial Training.}
In our experiments, we found that the mismatch between source and target data distribution could significantly degrade the performance of the mixed supervised learning framework. For example, if we choose the URBAN-TSD as the source dataset, and the L-TSD-weak as the target dataset, the performance will decrease substantially. This is because there is domain mismatch between URBAN-TSD and L-TSD-weak datasets \cite{pan2009survey,ben2006analysis}, i.e. \textit{f\_student} and \textit{w\_student} are first trained on the URBAN-TSD dataset but tested on L-TSD-weak dataset. To solve the domain mismatch problem, we propose a domain adversarial training strategy that aims to learn a common subspace shared by both the source and target domains, which enables all domains to have the same data distribution in the feature space. Specifically, inspired by GAN \cite{goodfellow2014generative} and DANN \cite{wang2019domain}, we make use of the adversarial relationship between modules Feature extractor (F) and Discriminator (D) to learn domain-invariant features in the feature space. To achieve this, we add an adversarial loss when we train \textit{f\_student} and \textit{w\_student}, as follows
\begin{equation}\label{ad loss}
    \mathcal{L}_{d}=||D(\boldsymbol{z})-\boldsymbol{d}||^2_{2}
\end{equation}
\begin{equation}\label{ad and sed}
    \mathcal{L}_{\mathit{d\_tsd}}=\mathcal{L}_{\mathit{tsd}}-\lambda_d*\mathcal{L}_{d}
\end{equation}
where $\boldsymbol{z}$ denotes the intermediate feature produced by F, and $\boldsymbol{d}$ denotes the domain label. The domain label is defined as $\boldsymbol{d}=[1,0]^T$ (which stands for the source domain) or $\boldsymbol{d}=[0,1]^T$ (target domain). $\mathcal{L}_{d}$ denotes the domain classification loss of the discriminator, $\mathcal{L}_{\mathit{tsd}}$ denotes the detection loss. $\mathcal{L}_{\mathit{d\_tsd}}$ denotes the training objective function, which minimizes the detection loss and meanwhile maximizes the domain classification loss. The parameter $\lambda_d$ controls the trade-off between $\mathcal{L}_{\mathit{tsd}}$ and $\mathcal{L}_{d}$. In our experiments, $\lambda_d$ is set to 0.2 empirically based on the validation set.\\
\textbf{Iterative Training Strategy.}
According to the previous description, we can use frame-level KD to transfer the knowledge from \textit{f\_student} to \textit{w\_student}, and use the pseudo supervised training to transfer the knowledge from \textit{w\_student} to \textit{f\_student}. Intuitively, the process can be repeated several times. Thus, an iterative training strategy is proposed. The whole algorithm is summarized in Algorithm 1.
\begin{algorithm}[htb]
\caption{Two-Student Learning}
\label{alg:Framwork}
\begin{algorithmic}[1] 
\REQUIRE ~~\\ 
    The source dataset $D_{s}$ and the target dataset $D_{t}$ \\
\ENSURE \textit{f\_student} and \textit{w\_student} model\\
    \STATE Training \textit{f\_student} on $D_{s}$ using formula (\ref{ssn}) and  (\ref{ad and sed})
    \STATE Training \textit{w\_student} on $D_{s}$ using formula (\ref{Mul-task loss}) and (\ref{ad and sed})
    \STATE Retraining \textit{w\_student} on $D_{t}$ using formula (\ref{wsn}) and (\ref{ad and sed})
    \STATE Retraining \textit{f\_student} on $D_{t}$ using formula (\ref{pseudo3}) and (\ref{ad and sed})
    \STATE While True: \\
           ~~ Retraining \textit{w\_student} on $D_{t}$, using formula (\ref{Mul-task loss}) \\
           ~~ Retraining \textit{f\_student} on $D_{t}$ using formula (\ref{pseudo3}) \\
           ~~ If no improvement: break;
\RETURN \textit{f\_student} and \textit{w\_student}; 
\end{algorithmic}
\end{algorithm}
\section{Experiments}
\subsection{Datasets}
In this section, we introduce the source dataset and target dataset used in our experiments. The source (\textit{resp.}, target) dataset is fully-annotated (\textit{resp.}, weakly-annotated).\\
\textbf{Source Dataset.} We first use the S-TSD dataset as the source dataset, which is a small-scale fully-annotated dataset based on Audioset \cite{hershey2021benefit}. In addition, we choose  10-category URBAN-TSD \cite{yang2021detect} as another source dataset, which includes two similar categories as in the L-TSD dataset: \textit{
dog\_bark} and \textit{gun\_shot}. \\
\textbf{Target Dataset.} We take the L-TSD-weak dataset as the target dataset. The details were given in Section \ref{audio-tsd}.

\subsection{Experimental Setups}
\noindent \textbf{Conditional Network.} We use the pre-trained PANNs \cite{kong2020panns} model to initialize the conditional network, and then fix it in the training process. \\
\textbf{Detection network.} All the raw audios are down-sampled to 22.05kHz and then Short Time Fourier Transform (STFT) with a window size of 2048 samples are applied, followed by a Mel-scaled filter bank on perceptually weighted spectrogram. This results in 64 Mel frequency bins and around 50 frames per second. When training \textit{f\_student} and \textit{w\_student} for the first time, the Adam optimizer \cite{kingma2015adam} is used for 100 epochs, with an initial learning rate of $1 \times 10^{-3}$. When they are re-trained, the learning rate is set as $1 \times 10^{-4}$. \\
\textbf{Metrics.} We use the segment-based F-score and event-based F-score \cite{mesaros2016metrics} as the evaluation metrics, which are the most commonly used metrics for detection task.
\begin{table}[t] \centering
\caption{F-score comparison with different supervision strategy on L-TSD test set. SS, WS and MS represent strong, weak and mixed supervision, respectively. F-scores are macro-averaged.}
\label{tab:my-table1}
\begin{tabular}{cccc}
\hline
Method              & Source dataset & Segment-F score  & Event-F score  \\ \hline
SS \cite{yang2021detect}                  &  -              & 58.57 & 50.4  \\ \cline{1-4}
WS \cite{yang2021detect}                  &  -              & 49.39 & 39.07 \\ \hline
\multirow{2}{*}{\textbf{MS (ours)}} & S-TSD          & 50.95 & 47.19 \\
                    & URBAN-TSD      & 51.31 & 47.56 \\ \hline
\end{tabular}
\end{table}
\subsection{Experiments on S-TSD and L-TSD Datasets}
In this section, we use S-TSD as the source dataset and L-TSD-weak as the target dataset. We compare our method with strongly supervised (SS) and weakly supervised (WS) methods. For SS-TSD, we directly train \textit{f\_student} on the L-TSD-strong dataset with the strong labels. For WS-TSD, we directly train \textit{w\_student} on the L-TSD-weak dataset with the weak labels. Note that for mixed supervised (MS) method, we only report the results obtained by \textit{f\_student} for the reason that \textit{f\_student} performs better than \textit{w\_student}. The experimental results are given in Table \ref{tab:my-table1}. From this table, we can see that our proposed MS method performs significantly better than the WS method, and performs similarly to the SS method. 

\subsection{Experiments on URBAN-TSD and L-TSD Datasets}
We also conduct experiments by using URBAN-TSD as the source dataset and L-TSD-weak as the target dataset. Table \ref{tab:my-table1} shows the experimental results, and we can see that using URBAN-TSD as the source dataset significantly improves the performance compared with the WS method. Furthermore, by comparing rows 3 and 4, we can find that using URBAN-TSD as the source dataset obtains better performance than using S-TSD as the source dataset. One of the reasons is the categories of URBAN-TSD and L-TSD-weak datasets have overlaps.
\begin{table}[t] \centering
\caption{Ablation studies on different strategies on L-TSD test set.}
\label{tab:my-table2}
\begin{tabular}{cccccc}
\hline
Model                                            & KD & PS & AD & Segment-F score   & Event-F score   \\ \hline
\multicolumn{1}{c|}{\multirow{3}{*}{w\_student}} & \usym{2613}    & \usym{2613}    & \usym{2613}   & 49.39 & 39.07 \\
\multicolumn{1}{c|}{}               & \usym{2613}  & \usym{2613}    & \checkmark   & 49.34 & 39.11 \\
\multicolumn{1}{c|}{}               & \checkmark  & \usym{2613}    & \usym{2613}   & 37.55 & 39.74 \\
\multicolumn{1}{c|}{}                            & \checkmark  & \usym{2613}   & \checkmark  & 50.34 & 41.35 \\ \hline
\multicolumn{1}{c|}{\multirow{3}{*}{f\_student}}  & \usym{2613}   & \checkmark  & \checkmark  & 48.09 & 43.47 \\
\multicolumn{1}{c|}{}                            & \checkmark  & \checkmark  & \usym{2613}   & 37.67 & 45.42 \\ 
\multicolumn{1}{c|}{}                            & \checkmark  & \checkmark  & \checkmark  & \textbf{51.31} & \textbf{47.56}  \\ \hline
\end{tabular}
\end{table}

\begin{table}[t] \centering
\caption{Ablation study on the effect of the number of iterations on iterative training strategy.}
\label{tab:my-table3}
\begin{tabular}{cccc}
\hline
Iterations                              & Model      & Segment-F score  & Event-F score  \\ \hline
\multicolumn{1}{c|}{\multirow{2}{*}{1}} & w\_student & 50.39 & 40.88 \\ \cline{2-4} 
\multicolumn{1}{c|}{}                   & f\_student & 50.21 & 46.99 \\ \hline
\multicolumn{1}{c|}{\multirow{2}{*}{2}} & w\_student & 50.14 & 41.60  \\ \cline{2-4} 
\multicolumn{1}{c|}{}                   & f\_student & 51.10  & 46.96 \\ \hline
\multicolumn{1}{c|}{\multirow{2}{*}{3}} & w\_student & 51.33 & 44.70  \\ \cline{2-4} 
\multicolumn{1}{c|}{}                   & f\_student & 50.95 & 47.19 \\ \hline
\end{tabular}
\end{table}
\subsection{Ablation Studies}
By taking URBAN-TSD as the source dataset, we conduct ablation studies to investigate the effectiveness of knowledge distillation (KD), pseudo supervised training (PS) and adversarial (AD) training, with the results shown in Table \ref{tab:my-table2}. For the \textit{w\_student} model: (1) The first row shows the results for directly training \textit{w\_student} on L-TSD-weak (without using any strategy). (2) The second row shows the results of directly training \textit{w\_student} on L-TSD-weak while using the AD strategy. We can see that only using the AD strategy does not give improvements, because it only aims to align the data distribution. (3) By comparing rows 2 and 4, we can see that the KD strategy can improve the \textit{w\_student}'s ability in capturing detailed information. (4) By comparing rows 3 and 4, we can see the effectiveness of the AD strategy when there is mismatch between the source and target datasets. 

For the \textit{f\_student} model: (1) By comparing rows 5 and 7, we can see that the KD strategy can improve the performance of \textit{f\_student} for the reason that it can improve the performance of \textit{w\_student}, and the better \textit{w\_student} leads to better \textit{f\_student}. (2) By comparing rows 6 and 7, we can find the improvements given by  the AD strategy. Lastly, we can see that \textit{f\_student} has better performance than \textit{w\_student}. A reason could be that the ability of \textit{f\_student} in capturing the detailed information is only partially transferred to \textit{w\_student} through the KD strategy. \\
\noindent \textbf{Influence of the number of iterations.} By taking S-TSD as the source dataset, we conduct ablation studies to investigate the influence of the number of iterations on the iterative training strategy, and the results are shown in Table \ref{tab:my-table3}. In this work, considering the cost of training time, we only tested three stages. For \textit{f\_student}, the segment- and event-based F-scores are increased from 50.21\% and 46.99\% to 50.95\% and 47.19\%, respectively. Furthermore, the event-based F-score of \textit{w\_student} is increased from 40.88\% to 44.7\%. It means that the two students can help each other.\\
\textbf{Why does pseudo supervised training work?}
To explain why the pseudo supervised training strategy is effective, we calculate the error rate between the true label and pseudo label (produced by \textit{w\_student}) on the L-TSD-strong training set. We found that the average error rate is 27.03\%, over the 490,336 audio clips. This is consistent with our empirical results as shown earlier in Figure \ref{fig:mask}: if the error rate is smaller than 0.35, the performance is similar to that of the case where the true label is taken as the target.
\section{Conclusions}
In this paper, we have presented a novel mixed supervised learning framework, which effectively improves the performance of novel categories with the help of a small-scale fully-annotated base categories dataset. In the future, we will apply our method to other tasks, such as SED and object detection. The source code and dataset of this work have been released\footnote{https://github.com/yangdongchao/weakly-target-sound-detection}.

\bibliographystyle{IEEEtran}
\bibliography{refs}

\begin{thebibliography}{10}
\providecommand{\url}[1]{#1}
\def\UrlFont{\rmfamily}
\providecommand{\newblock}{\relax}
\providecommand{\bibinfo}[2]{#2}
\providecommand\BIBentrySTDinterwordspacing{\spaceskip=0pt\relax}
\providecommand\BIBentryALTinterwordstretchfactor{4}
\providecommand\BIBentryALTinterwordspacing{\spaceskip=\fontdimen2\font plus
\BIBentryALTinterwordstretchfactor\fontdimen3\font minus
  \fontdimen4\font\relax}
\providecommand\BIBforeignlanguage[2]{{%
\expandafter\ifx\csname l@#1\endcsname\relax
\typeout{** WARNING: IEEEtran.bst: No hyphenation pattern has been}%
\typeout{** loaded for the language `#1'. Using the pattern for}%
\typeout{** the default language instead.}%
\else
\language=\csname l@#1\endcsname
\fi
#2}}

\bibitem{yang2021detect}
D.~Yang, H.~Wang, Y.~Zou, and C.~Weng, ``Detect what you want: Target sound
  detection,'' \emph{arXiv preprint arXiv:2112.10153}, 2021.

\bibitem{bello2018sonyc}
J.~P. Bello, C.~Silva, O.~Nov, R.~DuBois, A.~Arora, J.~Salamon, C.~Mydlarz, and
  H.~Doraiswamy, ``Sonyc: A system for the monitoring, analysis and mitigation
  of urban noise pollution,'' \emph{arXiv preprint arXiv:1805.00889}, 2018.

\bibitem{stowell2015acoustic}
D.~Stowell and D.~Clayton, ``Acoustic event detection for multiple overlapping
  similar sources,'' in \emph{IEEE Workshop on Applications of Signal
  Processing to Audio and Acoustics (WASPAA)}.\hskip 1em plus 0.5em minus
  0.4em\relax IEEE, 2015, pp. 1--5.

\bibitem{hershey2017cnn}
S.~Hershey, S.~Chaudhuri, D.~Ellis, J.~Gemmeke, A.~Jansen, R.~C. Moore,
  M.~Plakal, D.~Platt, R.~A. Saurous, B.~Seybold, \emph{et~al.}, ``{CNN}
  architectures for large-scale audio classification,'' in \emph{IEEE
  International Conference on Acoustics, Speech and Signal Processing
  (ICASSP)}.\hskip 1em plus 0.5em minus 0.4em\relax IEEE, 2017, pp. 131--135.

\bibitem{kong2019sound}
Q.~Kong, Y.~Xu, I.~Sobieraj, W.~Wang, and M.~D. Plumbley, ``Sound event
  detection and time-frequency segmentation from weakly labelled data,''
  \emph{IEEE/ACM Transactions on Audio, Speech, and Language Processing},
  vol.~27, pp. 777--787, 2019.

\bibitem{dinkel2021towards}
H.~Dinkel, M.~Wu, and K.~Yu, ``Towards duration robust weakly supervised sound
  event detection,'' \emph{IEEE/ACM Transactions on Audio, Speech, and Language
  Processing}, vol.~29, pp. 887--900, 2021.

\bibitem{lin2020specialized}
L.~Lin, X.~Wang, H.~Liu, and Y.~Qian, ``Specialized decision surface and
  disentangled feature for weakly-supervised polyphonic sound event
  detection,'' \emph{IEEE/ACM Transactions on Audio, Speech, and Language
  Processing}, vol.~28, pp. 1466--1478, 2020.

\bibitem{kong2020sound}
Q.~Kong, Y.~Xu, W.~Wang, and M.~D. Plumbley, ``Sound event detection of weakly
  labelled data with cnn-transformer and automatic threshold optimization,''
  \emph{IEEE/ACM Transactions on Audio, Speech, and Language Processing},
  vol.~28, pp. 2450--2460, 2020.

\bibitem{mesaros2021sound}
A.~Mesaros, T.~Heittola, T.~Virtanen, and M.~D. Plumbley, ``Sound event
  detection: A tutorial,'' \emph{IEEE Signal Processing Magazine}, vol.~38,
  no.~5, pp. 67--83, 2021.

\bibitem{wang2019comparison}
Y.~Wang, J.~Li, and F.~Metze, ``A comparison of five multiple instance learning
  pooling functions for sound event detection with weak labeling,'' in
  \emph{IEEE International Conference on Acoustics, Speech and Signal
  Processing (ICASSP)}.\hskip 1em plus 0.5em minus 0.4em\relax IEEE, 2019, pp.
  31--35.

\bibitem{martin2019sound}
I.~Mart{\'\i}n-Morat{\'o}, A.~Mesaros, T.~Heittola, T.~Virtanen, M.~Cobos, and
  F.~Ferri, ``Sound event envelope estimation in polyphonic mixtures,'' in
  \emph{IEEE International Conference on Acoustics, Speech and Signal
  Processing (ICASSP)}.\hskip 1em plus 0.5em minus 0.4em\relax IEEE, 2019, pp.
  935--939.

\bibitem{salamon2017scaper}
J.~Salamon, D.~MacConnell, M.~Cartwright, P.~Li, and J.~P. Bello, ``Scaper: A
  library for soundscape synthesis and augmentation,'' in \emph{IEEE Workshop
  on Applications of Signal Processing to Audio and Acoustics (WASPAA)}.\hskip
  1em plus 0.5em minus 0.4em\relax IEEE, 2017, pp. 344--348.

\bibitem{gemmeke2017audio}
J.~Gemmeke, D.~Ellis, D.~Freedman, A.~Jansen, W.~Lawrence, R.~C. Moore,
  M.~Plakal, and M.~Ritter, ``Audio set: An ontology and human-labeled dataset
  for audio events,'' in \emph{IEEE International Conference on Acoustics,
  Speech and Signal Processing (ICASSP)}.\hskip 1em plus 0.5em minus
  0.4em\relax IEEE, 2017, pp. 776--780.

\bibitem{hershey2021benefit}
S.~Hershey, D.~P. Ellis, E.~Fonseca, A.~Jansen, C.~Liu, R.~C. Moore, and
  M.~Plakal, ``The benefit of temporally-strong labels in audio event
  classification,'' in \emph{IEEE International Conference on Acoustics, Speech
  and Signal Processing (ICASSP)}.\hskip 1em plus 0.5em minus 0.4em\relax IEEE,
  2021, pp. 366--370.

\bibitem{kumar2017audio}
A.~Kumar and B.~Raj, ``Audio event and scene recognition: A unified approach
  using strongly and weakly labeled data,'' in \emph{International Joint
  Conference on Neural Networks (IJCNN)}.\hskip 1em plus 0.5em minus
  0.4em\relax IEEE, 2017, pp. 3475--3482.

\bibitem{liang2021joint}
Y.~Liang, Y.~Long, Y.~Li, and J.~Liang, ``Joint weakly supervised {AT} and
  {AED} using deep feature distillation and adaptive focal loss,'' \emph{arXiv
  preprint arXiv:2103.12388}, 2021.

\bibitem{lin2020guided}
L.~Lin, X.~Wang, H.~Liu, and Y.~Qian, ``Guided learning for weakly-labeled
  semi-supervised sound event detection,'' in \emph{IEEE International
  Conference on Acoustics, Speech and Signal Processing (ICASSP)}.\hskip 1em
  plus 0.5em minus 0.4em\relax IEEE, 2020, pp. 626--630.

\bibitem{shi2019hodgepodge}
Z.~Shi, L.~Liu, H.~Lin, R.~Liu, and A.~Shi, ``Hodgepodge: Sound event detection
  based on ensemble of semi-supervised learning methods,'' \emph{arXiv preprint
  arXiv:1907.07398}, 2019.

\bibitem{chan2020non}
T.~K. Chan, C.~S. Chin, and Y.~Li, ``Non-negative matrix
  factorization-convolutional neural network (nmf-cnn) for sound event
  detection,'' \emph{arXiv preprint arXiv:2001.07874}, 2020.

\bibitem{zheng2021improved}
X.~Zheng, Y.~Song, I.~McLoughlin, L.~Liu, and L.-R. Dai, ``An improved mean
  teacher based method for large scale weakly labeled semi-supervised sound
  event detection,'' in \emph{IEEE International Conference on Acoustics,
  Speech and Signal Processing (ICASSP)}.\hskip 1em plus 0.5em minus
  0.4em\relax IEEE, 2021, pp. 356--360.

\bibitem{cances2021comparison}
L.~Cances and T.~Pellegrini, ``Comparison of deep co-training and mean-teacher
  approaches for semi-supervised audio tagging,'' in \emph{IEEE International
  Conference on Acoustics, Speech and Signal Processing (ICASSP)}.\hskip 1em
  plus 0.5em minus 0.4em\relax IEEE, 2021, pp. 361--365.

\bibitem{guan2022sparse}
Y.~Guan, J.~Xue, G.~Zheng, and J.~Han, ``Sparse self-attention for
  semi-supervised sound event detection,'' in \emph{IEEE International
  Conference on Acoustics, Speech and Signal Processing (ICASSP)}.\hskip 1em
  plus 0.5em minus 0.4em\relax IEEE, 2022, pp. 821--825.

\bibitem{kong2020panns}
Q.~Kong, Y.~Cao, T.~Iqbal, Y.~Wang, W.~Wang, and M.~Plumbley, ``Panns:
  Large-scale pretrained audio neural networks for audio pattern recognition,''
  \emph{IEEE/ACM Transactions on Audio, Speech, and Language Processing},
  vol.~28, pp. 2880--2894, 2020.

\bibitem{pan2009survey}
S.~J. Pan and Q.~Yang, ``A survey on transfer learning,'' \emph{IEEE
  Transactions on Knowledge and Data Engineering}, vol.~22, no.~10, pp.
  1345--1359, 2009.

\bibitem{ben2006analysis}
S.~Ben-David, J.~Blitzer, K.~Crammer, and F.~Pereira, ``Analysis of
  representations for domain adaptation,'' \emph{Advances in Neural Information
  Processing Systems}, vol.~19, 2006.

\bibitem{goodfellow2014generative}
I.~J. Goodfellow, J.~Pouget-Abadie, M.~Mirza, B.~Xu, D.~Warde-Farley, S.~Ozair,
  A.~Courville, and Y.~Bengio, ``Generative adversarial networks,'' \emph{arXiv
  preprint arXiv:1406.2661}, 2014.

\bibitem{wang2019domain}
R.~Wang, M.~Wang, X.-L. Zhang, and S.~Rahardja, ``Domain adaptation neural
  network for acoustic scene classification in mismatched conditions,'' in
  \emph{Asia-Pacific Signal and Information Processing Association Annual
  Summit and Conference (APSIPA ASC)}.\hskip 1em plus 0.5em minus 0.4em\relax
  IEEE, 2019, pp. 1501--1505.

\bibitem{kingma2015adam}
D.~Kingma and J.~Ba, ``Adam: A method for stochastic optimization,'' in
  \emph{ICLR}, 2015.

\bibitem{mesaros2016metrics}
A.~Mesaros, T.~Heittola, and T.~Virtanen, ``Metrics for polyphonic sound event
  detection,'' \emph{Applied Sciences}, vol.~6, no.~6, p. 162, 2016.

\end{thebibliography}

%
%
%
%
%
%
%
%
%

\end{sloppy}
\end{document}